\documentclass[preprint,12pt]{elsarticle}

\usepackage{amssymb}

\usepackage{float}
\usepackage{xcolor}
\usepackage{hyperref}
\usepackage{fullpage}

\restylefloat{table}

\journal{SoftwareX}

\begin{document}

\begin{frontmatter}



\title{UQpy v4.1: Uncertainty Quantification with Python}


\author{Dimitrios Tsapetis$^{1*}$}
\author{Michael D. Shields$^1$}
\author{Dimitris G. Giovanis$^1$}
\author{Audrey Olivier$^2$}
\author{Lukas Novak$^3$}
\author{Promit Chakroborty$^1$}
\author{Himanshu Sharma$^1$}
\author{Mohit Chauhan$^1$}
\author{Katiana Kontolati$^1$}
\author{Lohit Vandanapu$^1$}
\author{Dimitrios Loukrezis$^4$}
\author{Michael Gardner$^5$}

\address{$^1$Department of Civil and Systems Engineering, Johns Hopkins University, Baltimore, MD, 21218, USA}
\address{$^2$Sonny Astani Department of Civil and Environmental Engineering, University of Southern California, Los Angeles, CA, 90089}
\address{$^3$Faculty of Civil Engineering, Brno University of Technology, Veveri 331/95, Brno 60200, Czech Republic}
\address{$^4$Department of Electrical Engineering and Information Technology, Technical University of Darmstadt, Darmstadt, Germany}
\address{$^5$Department of Civil and Environmental Engineering, University of California, Davis, CA, 95616}

\begin{abstract}
This paper presents the latest improvements introduced in v4.1 of the UQpy, Uncertainty Quantification with Python, library. In the latest version, the code was restructured to conform with the latest Python coding conventions, refactored to simplify previous tightly coupled features, and improve its extensibility and modularity. To improve the robustness of UQpy, software engineering best practices were adopted. A new software development workflow significantly improved collaboration between team members, and continous integration and automated testing ensured the robustness and reliability of software performance. Continuous deployment of UQpy allowed its automated packaging and distribution in system agnostic format via multiple channels, while a Docker image enables the use of the toolbox regardless of operating system limitations.

\end{abstract}

\begin{keyword}
Uncertainty Quantification \sep Continuous Integration 



\end{keyword}

\end{frontmatter}

\section*{Required Metadata}
\label{sec:RequiredMetadata}

\section*{Current code version}
\label{sec:CurrentVersion}

\begin{table}[H]
\begin{tabular}{|l|p{6.5cm}|p{6.5cm}|}
\hline
\textbf{Nr.} & \textbf{Code metadata description} & \textbf{Please fill in this column} \\
\hline
C1 & Current code version & v4.1.1 \\
\hline
C2 & Permanent link to code/repository used for this code version & \texttt{https://github.com/SURGroup/ UQpy}\\
\hline
C3 & Code Ocean compute capsule &  \texttt{https://codeocean.com/ capsule/9924279/tree} \\
\hline
C4 & Legal Code License   & MIT Licence \\
\hline
C5 & Code versioning system used & git \\
\hline
C6 & Software code languages, tools, and services used & Python, MPI \\
\hline
C7 & Compilation requirements, operating environments \& dependencies & see \texttt{setup.py}\\
\hline
C8 & If available Link to developer documentation/manual &  \texttt{https://uqpyproject. readthedocs.io} \\
\hline
C9 & Support forum for questions & \texttt{https://github.com/SURGroup/ UQpy/discussions} \\ 
\hline
\end{tabular}
\caption{Code metadata}
\label{tab:code_metadata} 
\end{table}


\section{Motivation and significance} \label{sec:Intro}

Uncertainty Quantification (UQ) is the science of characterizing, quantifying, managing, and reducing uncertainties in mathematical, computational and physical systems. Depending on the sources of uncertainty, UQ provides a multitude of methodologies to quantify their effects. For instance, given the probability distribution for the inputs to a computational model, forward uncertainty propagation methods aim to estimate the distributions or statistics of resulting quantities of interest. Inverse UQ, on the other hand, aims to infer uncertainties in input quantities given limitations and uncertainties in the observed system response, e.g.\ for model calibration from experimental data. Numerous related tasks fall under the broad classification of UQ including sensitivity analysis, which aims to quantify the influence of multiple inputs to a system, and reliability analysis which aims to estimate (and sometimes minimize) the probability of failure of the system.




A major challenge in UQ is to reduce the high computational expense associated with many repeated model evaluations. This can be achieved through advances in sampling, development of computationally inexpensive surrogate models (or metamodels), and by leveraging high performance computing. To address these challenges, multiple software packages and libraries have been developed. Some of the most comprehensive libraries for UQ include \texttt{OpenTurns}\cite{OpenTurns}, \texttt{Korali}\cite{Korali}, \texttt{MUQ}\cite{MUQ}, \texttt{UQTk}\cite{UQTk}, \texttt{Dakota}\cite{Dakota}, \texttt{OpenCossan}\cite{OpenCossan} and \texttt{UQLab}\cite{UQLab}. These software are developed in either \texttt{C}, \texttt{C++} programming languages or \texttt{Matlab} and although many provide bindings to \texttt{Python} (to differing extents), they are not generally suitable for direct extension in \texttt{Python}, which is one of the most widely used languages in the scientific community. 

Apart from these general purpose UQ libraries, several packages that target specific applications or with more limited scope are available. In \texttt{R}, the \texttt{DiceDesign}\cite{DiceDesign} package aids experimental design, while \texttt{DiceKriging} and \texttt{DiceOptim}\cite{DiceKriging} use Kriging for metamodeling and surrogate-based optimization, respectively.  The \texttt{Matlab} code \texttt{FERUM}\cite{FERUM}, developed by the Engineering Risk Analysis Group at the Technical University of Munich, serves as a general purpose finite element structural reliability code, while \texttt{SUMOToolbox}\cite{SumoToolbox} is a framework for global surrogate modelling and adaptive sampling. Specifically in Python, several focused libraries have been developed.
\texttt{UncertaintyPy}\cite{UncertaintyPy} was developed for UQ in computational neuroscience. \texttt{PyROM} framework\cite{PyROM} provides a user-friendly way to implement model reduction techniques. The \texttt{ChaosPy} package provides UQ functionality centered around polynomial chaos expansions. Bayesian calibration algorithms are implemented in \texttt{SPUX}\cite{SPUX} and \texttt{ABCpy}\cite{ABCpy} and sensitivity analyses by \texttt{SALib}\cite{SALib}. \texttt{PyMC}\cite{PyMC}  provides a simple Python interface that allows its user to create Bayesian models and fit them using Markov Chain Monte Carlo methods. \texttt{PyGPC}\cite{PyGPC} library is based on generalized polynomial chaos theory and provides capabilities for uncertainty and sensitivity analysis of computational models. Three of the latest additions are \texttt{PyApprox}\cite{PyApprox}, which provides wide-ranging functionality, \texttt{NeuralUQ}\cite{NeuralUQ} focused on UQ in neural network models, and \texttt{Fortuna} that provides uncertainty estimates, classification and prediction for production systems. 

\texttt{UQpy} aims to provide a comprehensive UQ library with wide-ranging capabilities spanning the areas discussed above, as well as a development environment for creating new UQ methodologies. The \texttt{UQpy} package was originally introduced in \cite{UQpyv3}, where the overall structure of v3 was described. Since then, the authors have reworked the \texttt{UQpy} architecture with the goal to simplify its structure, enhance its extensibility, and make it more robust. The updated architecture of the library rendered it not backwards compatible, as the strategy for construction of classes has changed. Yet porting older solutions to the new structure can be performed in a straightforward manner. This restructuring resulted in the current version we present here, v4.1.

The first task carried out towards v4.1 was to restructure the file system. The previous structure which maintained a single \texttt{Python} file per module had reached size limitations and made it cumbersome for the team members to add new functionalities or update existing ones. In the reorganization, a directory was created for each module, which contains, in a hierarchical structure, subdirectories for specific functionalities, with one file dedicated to each class. Slight modifications were also made to the existing code to ensure compliance with \texttt{PEP8} by renaming modules, classes, and function signatures. Instead of monolithic classes per functionality, each component was split into a separate class with a dedicated \texttt{abstract baseclass}, where applicable. This choice reduced code complexity, provided a standardized way of extending components, and enabled the construction of the final functionality, using object composition and inheritance.

The second step to improve team collaboration was to deprecate the ``branch-per-developer'' strategy and move to a feature-based branch structure using the Github Flow. This removed unnecessary redundancies and complications when multiple people are working on related functionalities. At the same time, the workflow is now directly combined with testing automation and Continuous Integration/Continuous Delivery (CI/CD) workflows. Unit tests were implemented throughout the software, achieving code coverage greater than 80\%. The CI pipeline includes linting, code quality checks, and automated semantic versioning, while the CD pipeline packages and distributes the code via multiple channels, such as \texttt{PyPI}, \texttt{conda-forge}, and \texttt{Docker} images. This CI/CD pipelines are explained in more detail in Section \ref{sec:ContinuousIntegration}. 

The documentation was revamped to reflect the new hierarchical structure of the code, with embedded examples serving as tutorials to quickly familiarize users with the code functionality. Specifically, for each class, a gallery of examples is created using the \texttt{sphinx-gallery} extension \cite{sphinx_gallery}. The users can now download the examples in both Jupyter notebook and Python format  or directly interact with the example in a dedicated Binder environment. Finally, several new functionalities were introduced either by the development team or external collaborations, thus boosting \texttt{UQpy}'s capabilities. 

\section{Software description}
\label{sec:SoftwareDescription}

\subsection{Software Architecture}
\label{sec:SoftwareArchitecture}

\texttt{UQpy} is a Python-based toolbox that provides a series of computational methodologies and algorithms for wide-ranging UQ problems.  The core of \texttt{UQpy} is based on state-of-the-art Python libraries, specifically \texttt{NumPy} \cite{Numpy}, which is the most fundamental package supporting array and linear algebra operations, \texttt{SciPy} \cite{scipy}, that provides algorithms for optimization, integration and basic statistics, and \texttt{scikit-learn} \cite{scikit_learn}, which includes various tools for supervised and unsupervised learning. \texttt{UQpy} is split into eleven modules, nine of which address specific tasks in UQ and which will be discussed in detail in the following section. A module that enables necessary simulations in all other modules, called \texttt{run\_model}, aids in the batch execution of both \texttt{Python} and third-party computational models and includes functionality for parallelization via MPI for high performance computing. Finally, a \texttt{utilities} module contains various functions that are common to multiple modules.

\subsection{Software Modules}
\label{sec:software_functionalities}
In this section, all existing modules of \texttt{UQpy} will be briefly introduced, with emphasis on software updates compared to v3. 
The respective \texttt{UML} diagrams are included in the \texttt{UQpy} documentation allowing architecture visualization.

\subsubsection{\texttt{distributions} module}
The \texttt{distributions} module serves as the basis for most probabilistic operations in \texttt{UQpy}. It is fully compatible with \texttt{scipy} distributions and enables users to create probabilistic distribution objects. Compared to the previous version, the baseclass hierarchy was simplified. An abstract baseclass \texttt{Distribution} serves as the interface for creating all subsequent distributions. Depending on the dimensionality of the distribution this baseclass is further refined into \texttt{Distribution1D} and \texttt{DistributionND} for univariate and multivariate distributions respectively, while the \texttt{Distrubution1D} is futher subclassed into \texttt{DistributionsContinuous1D} and \texttt{DistributionsDiscrete1D} for continuous and discrete random variables. Within this structure, 23 distinct distributions are implemented. A \texttt{Copula} baseclass with two implementations enables users to add dependence between 1D distribution objects. All baseclasses can be easily extended by users to implement any distribution of their choice by simply creating a new child class for the distribution and implementing the requisite methods.

\subsubsection{\texttt{sampling} module}
This module provides a wide range of methods to draw samples of random variables. The following classes enable Monte Carlo simulation and variance reduction methods: \texttt{MonteCarloSampling}, \texttt{SimplexSampling}, \texttt{ImportanceSampling}, and \texttt{StratifiedSampling}. The \texttt{StratifiedSampling} class has been refactored as a parent class for all stratified sampling approaches with \texttt{LatinHypercubeSampling}, \texttt{TrueStratifiedSampling}, and \texttt{RefinedStratifiedSampling} as child classes, all of which utilize a common \texttt{Strata} class for geometric decomposition of the domain. Markov-Chain Monte Carlo (MCMC) methods are included, with the \texttt{MCMC} abstract baseclass serving as the common interface and 7 different methodologies implemented as subclasses. The latest version includes two new implementations of parallel and sequential tempering MCMC algorithms. Additional MCMC methods can be implemented by the user by simply creating a new subclass with the requisite methods.
The module also includes the \texttt{AdaptiveKriging} for adaptive sample generation for Gaussian process surrogate modeling (see Section \ref{sec:surrogates}) using specified (and custom) learning functions. Compared to v3, all learning functions have been extracted as separate classes, with a common \texttt{LearningFunction} baseclass, allowing users to easily create custom implementations.

\subsubsection{\texttt{transformations} module}
This module contains isoprobabilistic transformations of random variables. Except for updates in naming conventions, this module retained the previous functionality with the \texttt{Nataf}, \texttt{Correlate}, and \texttt{Decorelate} transformations being available.

\subsubsection{\texttt{stochastic\_process} module}
This module supports the simulation of univariate, multivariate, and multidimensional Gaussian and non-Gaussian stochastic processes, with the latest addition since v3 being the two-dimensional Karhunen-Lo\`eve Expansion in the \texttt{KarhunenLoeveExpansion2D} class. All pre-existing classes of \texttt{SpectralRepresentation}, \texttt{BispectralRepresentation} and \texttt{KarhunenLoeveExpansion} have been updated to conform with \texttt{PEP8} Python coding standards.

\subsubsection{\texttt{run\_model} module}
This module is not directly related to any specific UQ operations, yet it is an integral part of the \texttt{UQpy} software. It lies at its core and supports the execution of either Python or third-party computational models at specified sampling points.

\texttt{UQpy} interfaces Python models directly, by importing and executing the code. On the other hand, \texttt{UQpy} interfaces with third-party software models through ASCII text files to introduce uncertainties in their inputs and uses a standardized scripting format for model execution. In both cases, \texttt{UQpy} supports serial and parallel execution. Parallel execution allows the execution of different samples simultaneously, with options for local and cluster execution. Local parallel execution uses MPI and the \texttt{mpi4py} library to distribute the random samples among tasks that are processed independently. In this case, the model evaluation cannot invoke MPI internally. In cluster enabled parallelization, with the aid of a bash script, a tiling of the jobs can be performed to include both shared and distributed memory parallelism, while enabling the user to work with different HPC schedulers.

\subsubsection{\texttt{dimension\_reduction} module}
In the update from v3 to v4.1, the \texttt{dimension\_reduction} module was rewritten from scratch. The existing \texttt{DirectPOD} and \texttt{SnapshotPOD} methods were reworked to comply with the latest Python coding conventions and the \texttt{HigherOrderSVD} class was added. To support Grassmann manifold projections and operations, a series of classes were added. The \texttt{GrassmannProjection} class serves as the parent for classes that project data arrays onto the manifold, with the \texttt{SVDprojection} subclass currently available. After the data have been projected, operations such as computing the Karcher mean or Frechet variance are available with the aid of the \texttt{GrassmannOperations} class. Interpolation can be performed on the manifold with the \texttt{GrassmannInterpolation} class. Special attention was given to the \texttt{DiffusionMaps} class, where the kernel computation was extracted and delegated to a hierarchy of kernel classes in the \texttt{utilities} modules for broader use in future development of other kernel-based methods. More detail can be found in Section \ref{sec:utilities}.

\subsubsection{\texttt{inference} module}
The functionality of the \texttt{inference} module was retained from v3 to v4.1 but restructured. The previous  \texttt{InferenceModel} class, which defines the model on which inference is performed, has now been split into three separate classes depending on the specific model type, namely \texttt{DistributionModel}, \texttt{LogLikelihoodModel} and \texttt{ComputationalModel}, all under the revised \texttt{InferenceModel} baseclass. For information theoretic-based model selection using the \texttt{InformationModelSelection} class, the information criteria have been extracted as separate classes, \texttt{AIC}, \texttt{BIC}, \texttt{AICc}, under a new common \texttt{InformationCriterion} baseclass. The remaining functionality of \texttt{MLE}, \texttt{BayesParameterEstimation}, and \texttt{BayesModelSelection} was updated according to the newly adopted coding conventions and, for Bayesian evidence computation, the \texttt{EvidenceMethod} baseclass has been established with the \texttt{HarmonicMean} subclass defined and allowing straightforward implementation of new Bayesian evidence methods as distinct subclasses.

\subsubsection{\texttt{reliability} module}

Modifications to the \texttt{reliability} module were made to ensure compliance with the latest Python coding guidelines. The first and second-order reliability methods, \texttt{FORM} and \texttt{SORM}, were restructured as subclasses under a common \texttt{TaylorSeries} baseclass to remove code redundancies. The existing \texttt{SubsetSimulation} class was retained and revised to match best practices.

\subsubsection{\texttt{surrogates} module}
\label{sec:surrogates}
One of the most heavily refactored modules in the latest version is \texttt{surrogates}. Generally, surrogate models are now developed under the abstract \texttt{Surrogate} baseclass. The previously existing \texttt{Kriging} class was removed entirely and is now replaced with the more general \texttt{GaussianProcessRegression}, which includes the functionality to perform regression or interpolation (Kriging). Kernels are extracted as separate classes, with the abstract baseclass \texttt{Kernel} (from the \texttt{utilities} module), serving as an interface. The \texttt{RBF} and \texttt{Matern} kernels have been implemented. For use with \texttt{GaussianProcessRegression}, multiple regression methods are implemented as subclasses under the \texttt{Regression} baseclass. The newest addition to \texttt{GaussianProcessRegression} is the ability to add constraints using the virtual point method. These constraints are implemented under the \texttt{Constraints} baseclass, which makes adding new constraints straightforward by implementing a new subclass with the requisite methods. 

The \texttt{PolynomialChaosExpansion} class was rewritten from scratch (now as a subclass of \texttt{Surrogate}) to resolve performance issues. Two new baseclasses were introduced. The \texttt{Polynomials} baseclass defines sets of orthogonal polynomials as subclasses, including the \texttt{Hermite} and \texttt{Legendre} polynomial classes. The \texttt{PolynomialBasis} baseclass establishes a set of subclasses to define the polynomial basis, e.g.\ using a classical tensor product basis \texttt{TensorProductBasis}, or introducing new ways to reduce the basis computation such as the \texttt{TotalDegreeBasis} and \texttt{HyperbolicBasis} classes. This makes the code easily extensible to include new means of basis construction. All regression methods were united as subclasses under the \texttt{Regression} baseclass, again making it more easily extended for new methods, and the computationally efficient \texttt{LeastAngleRegression} was added. 

Lastly, the \texttt{SROM} method was retained and updated to conform with the latest Python software development practices.

\subsubsection{\texttt{sensitivity} module}

In v3, the \texttt{sensitivity} module only contained the \texttt{MorrisSensitivity} method. This module significantly benefited from the extensibility introduced in \texttt{UQpy} with v4.1. The \texttt{Sensitivity} abstract baseclass now contains the first major contribution from external collaborators introduced in a set of subclasses that include \texttt{SobolSensitivity}, \texttt{GeneralizedSobolSensitivity}, \texttt{ChatterjeeSensitivity}, and \texttt{CramerVonMisesSensitivity}. Additionally, the updated polynomial chaos expansion code in the \texttt{surrogates} module (see Section \ref{sec:surrogates}), allows the computation of first and total order sensitivity indices with reduced computational cost through the \texttt{PceSensitivity} class, which takes advantage of a fitted \texttt{PolynomialChaosExpansion} object.

\subsubsection{\texttt{utilities} module}
\label{sec:utilities}

The new \texttt{utilities} module contains code that may be used in multiple modules. This currently contains two abstract baseclasses, the \texttt{Kernel} baseclass and the \texttt{Distance} baseclass for computing kernels and measures of distance, respectively. Within each baseclass, there are two additional baseclasses for Euclidean and Grassmannian kernels/distances. Several kernels and distances have been added as subclasses and new ones can be easily developed by writing a new subclass with the requisite methods.

\section{Continuous Integration}\label{sec:ContinuousIntegration}

\texttt{UQpy} v3\cite{UQpyv3} was developed using the flexible standards of an academic software, which challenged the ability of the team to collaborate and develop new features using a streamlined workflow. To this end, the latest version was fully restructured to enhance its extensibility, while modern software development practices were introduced to support collaboration and ensure code robustness and quality. The standard of Github Flow was adopted as development strategy. The \texttt{master} branch of the Github repository always contains the latest stable version. A \texttt{Development} branch is now used for merging all newly developed functionality and bug fixes. New versions of the software are released when a pull request is merged from the \texttt{Development} branch to \texttt{master}. For developing new features, a \texttt{feature-\{functionality\}} branch is created from the latest \texttt{Development} state and merged back once complete. The case is similar for bug fixes, with branches following the \texttt{bugfix-\{bug\}} naming convention. The aforementioned workflow enables a consistent way of treating new functionality or addressing errors arising during development. 

To ensure the code quality of all previously implemented features, the development team enforced unit testing practices. Since the functionality implemented in UQpy is inherently stochastic, and its randomness stems from random number generators, a process of setting the seed to ensure test reproducibility is adopted.  All previous functionalities are tested against benchmark problems to achieve a minimum of $80 \%$ line coverage. To ensure that the code coverage directive is enforced, Azure Pipelines were used to automatically run all tests and compute coverage when a commit is pushed to the Github repository. The static code analyzer \texttt{Pylint} is also used to enforce coding standards and ensure that no syntax errors are allowed. In addition to these checks, a code quality tool named \texttt{SonarCloud} is used to eliminate code vulnerabilities. This tool is triggered when creating a pull request and automatically detects any code smells, bugs or code duplications introduced, and fails when exceeding a predefined threshold. For a pull request to be acceptable, all test, linting, and code quality must satisfy minimum acceptance criteria and must pass a detailed code review from the code owners. Only then will the additions be merged to \texttt{Development} and subsequently to \texttt{master} branch.

Apart from the Continuous Integration process mentioned above, that ensure the robustness of \texttt{UQpy}, a set of Continuous Deployment (CD) actions are triggered. The first action is to evoke the \texttt{GitVersion} tool, which traverses the Git history of the code and determines the version of the code automatically, as a sequence of numbers \texttt{v\{Major\}.\{Minor\}.\{Patch\}}. Using the computed version, the code is packaged and automatically distributed to Python Packaging Index (\texttt{PyPI}), Github release inside the repository, as well as a \texttt{Docker} image that contains the latest \texttt{UQpy} version.

Finally, a structured logging framework was established -- in lieu of \texttt{print} commands triggered by \texttt{if} statements that were previously used to indicate errors or faults -- that allows users to select the required level of severity tracked during code execution. Six different levels of severity are available in \texttt{Python}, namely \texttt{NOTSET}, \texttt{DEBUG}, \texttt{INFO}, \texttt{WARN}, \texttt{ERROR}, and \texttt{CRITICAL}, with the \texttt{ERROR} being the default case in \texttt{UQpy}. The users can choose a more verbose setting by opting for the \texttt{INFO} severity level. Logging output is then directed to their sinks of choice e.g. Terminal, Logfile, Http Streams, etc. 

\section{Impact} \label{sec:Impact}

The latest version of \texttt{UQpy} modernizes the software to meet best practices in scientific software development, while also updating and improving functionality. This makes the package easier to use and more robust, broadens the classes of problems that it can solve, and greatly enhances the developement experience. These points are critical to the widespread adoption of UQ in scientific applications. This robust yet friendly Python library is both user- and developer-friendly and provides core functionality to casual users, state-of-the-art methods for advanced users, and a carefully designed environment for developers of UQ methods. With the advent of version 4, we have seen the user-base increase as the library has been adopted by external UQ teams, and have now successfully integrated updates from third-party developers -- both of which serve to advance the field of UQ. 

To summarize, the entire package has been restructured from a single-file per area to a module hierarchy. Wherever possible, suboptions inside algorithms were extracted using the  Strategy design pattern to enhance encapsulation and allow users to select their functionality in a more clear and straightforward manner. Baseclasses are now used throughout the code, which provides interfaces for the implementation of new algorithmic alternatives. To enhance the team collaboration efforts, the already existing version control and Github repository were supported with a CI/CD pipeline that automates software testing and code quality checks to ensure the best scientific output, while each new merge to the master is followed by package releases to \texttt{PyPI}, \texttt{conda-forge}, and Dockerhub image repository.

Compared to the other existing UQ packages, many of which have been listed above, the aim of \texttt{UQpy} is twofold. First of all, we aim to provide an extensive UQ library that addresses the wide-ranging needs of the scientific community. At the same time, we want to provide a toolbox that allows its straightforward extension with new functionalities and its use in real-world UQ applications. The developments outlined here represent significant advancements toward these two objectives.

\section{Conclusions}\label{sec:conclusions}
In this work, the open-source library for uncertainty quantification \texttt{UQpy} and specifically the latest v4.1 was introduced. All changes and updates to the modules of the library were explained in detail, with one of the most significant being the new software development and continuous integration workflow. The latest version enables users and external collaborators to expedite the development of new features using \texttt{UQpy} as a platform. This is proven by the new functionalities introduced from both the development team, as well as external collaborators.

\section{Conflict of Interest}
We confirm that there are no known conflicts of interest associated with this publication and there has been no significant financial support for this work that could have influenced its outcome.

\section*{Acknowledgements}
\label{sec:acknoledgements}
The authors gratefully acknowledge the following individuals who contributed to code development in \texttt{UQpy}: Ulrich Römer, TU Braunschweig; Julius Schultz, TU Braunschweig; Prateek Bhustali, TU Braunschweig. The authors of this work have been supported by the following sources: U.S. Dept. of Energy Award Number DE-SC0020428 (DT, MDS, DG, KK) and Defense Threat Reduction Agency Award Number DTRA1-20-2-0001 (DT, MDS, DG, HS); INL Laboratory Directed Research \& Development (LDRD) Program under DOE Idaho Operations Office Contract DE-AC07-05ID14517 (PC);  National Science Foundation under award numbers 1652044 (LV) and 132278 (MC). The content of the information does not necessarily reflect
the position or the policy of the federal government, and no official
endorsement should be inferred.




\begin{thebibliography}{00}



\bibitem{Numpy}Harris, C., Millman, K., Walt, S., Gommers, R., Virtanen, P., Cournapeau, D., Wieser, E., Taylor, J., Berg, S., Smith, N., Kern, R., Picus, M., Hoyer, S., Kerkwijk, M., Brett, M., Haldane, A., Río, J., Wiebe, M., Peterson, P., Gérard-Marchant, P., Sheppard, K., Reddy, T., Weckesser, W., Abbasi, H., Gohlke, C. \& Oliphant, T. Array programming with NumPy. {\em Nature}. \textbf{585}, 357-362 (2020,9), https://doi.org/10.1038/s41586-020-2649-2

\bibitem{scipy}Virtanen, P., Gommers, R., Oliphant, T., Haberland, M., Reddy, T., Cournapeau, D., Burovski, E., Peterson, P., Weckesser, W., Bright, J., Van der Walt, S., Brett, M., Wilson, J., Millman, K., Mayorov, N., Nelson, A., Jones, E., Kern, R., Larson, E., Carey, C., Polat, İ., Feng, Y., Moore, E., VanderPlas, J., Laxalde, D., Perktold, J., Cimrman, R., Henriksen, I., Quintero, E., Harris, C., Archibald, A., Ribeiro, A., Pedregosa, F., Van Mulbregt, P. \& SciPy 1.0 Contributors SciPy 1.0: Fundamental Algorithms for Scientific Computing in Python. {\em Nature Methods}. \textbf{17} pp. 261-272 (2020)

\bibitem{scikit_learn}Pedregosa, F., Varoquaux, G., Gramfort, A., Michel, V., Thirion, B., Grisel, O., Blondel, M., Prettenhofer, P., Weiss, R., Dubourg, V., Vanderplas, J., Passos, A., Cournapeau, D., Brucher, M., Perrot, M. \& Duchesnay, E. Scikit-learn: Machine Learning in Python. {\em Journal Of Machine Learning Research}. \textbf{12} pp. 2825-2830 (2011)

\bibitem{UQpyv3}Olivier, A., Giovanis, D., Aakash, B., Chauhan, M., Vandanapu, L. \& Shields, M. UQpy: A general purpose Python package and development environment for uncertainty quantification. {\em Journal Of Computational Science}. \textbf{47} pp. 101204 (2020), https://www.sciencedirect.com/science/article/pii/S1877750320305056

\bibitem{OpenTurns}Baudin, M., Dutfoy, A., Iooss, B. \& Popelin, A. Open TURNS: An industrial software for uncertainty quantification in simulation. (arXiv,2015), https://arxiv.org/abs/1501.05242

\bibitem{Korali}Martin, S., Wälchli, D., Arampatzis, G., Economides, A., Karnakov, P. \& Koumoutsakos, P. Korali: Efficient and scalable software framework for Bayesian uncertainty quantification and stochastic optimization. {\em Computer Methods In Applied Mechanics And Engineering}. \textbf{389} pp. 114264 (2022), https://www.sciencedirect.com/science/article/pii/S0045782521005752

\bibitem{MUQ}Parno, M., Davis, A. \& Seelinger, L. MUQ: The MIT Uncertainty Quantification Library. {\em Journal Of Open Source Software}. \textbf{6}, 3076 (2021), https://doi.org/10.21105/joss.03076

\bibitem{UQTk}Debusschere, B., Najm, H., Pébay, P., Knio, O., Ghanem, R. \& Le Maıtre, O. Numerical Challenges in the Use of Polynomial Chaos Representations for Stochastic Processes. {\em SIAM Journal On Scientific Computing}. \textbf{26}, 698-719 (2004)

\bibitem{Dakota}Dalbey, K., Eldred, M., Geraci, G., Jakeman, J., Maupin, K., Monschke, J., Seidl, D., Swiler, L., Tran, A., Menhorn, F. \& Zeng, X. Dakota A Multilevel Parallel Object-Oriented Framework for Design Optimization Parameter Estimation Uncertainty Quantification and Sensitivity Analysis: Version 6.12 Theory Manual.  (2020,5), https://www.osti.gov/biblio/1630693

\bibitem{OpenCossan}Patelli, E., Broggi, M., Angelis, M. \& Beer, M. OpenCossan: An Efficient Open Tool for Dealing with Epistemic and Aleatory Uncertainties.  (2014,7)

\bibitem{UQLab}Marelli, S. \& Bruno Sudret UQLab: A Framework for Uncertainty Quantification in Matlab. {\em Vulnerability, Uncertainty, And Risk}. pp. 2554-2563, https://ascelibrary.org/doi/abs/10.1061/9780784413609.257

\bibitem{DiceDesign}Dupuy, D., Helbert, C. \& Franco, J. DiceDesign and DiceEval: Two R Packages for Design and Analysis of Computer Experiments. {\em Journal Of Statistical Software}. \textbf{65}, 1-38 (2015), https://www.jstatsoft.org/index.php/jss/article/view/v065i11

\bibitem{DiceKriging}Roustant, O., Ginsbourger, D. \& Deville, Y. DiceKriging, DiceOptim: Two R Packages for the Analysis of Computer Experiments by Kriging-Based Metamodeling and Optimization. {\em Journal Of Statistical Software}. \textbf{51}, 1-55 (2012), https://www.jstatsoft.org/index.php/jss/article/view/v051i01

\bibitem{TUM}Munich, E. Software. (https://www.cee.ed.tum.de/era/software/,2023)

\bibitem{SumoToolbox}Gorissen, D., Couckuyt, I., Demeester, P., Dhaene, T. \& Crombecq, K. A Surrogate Modeling and Adaptive Sampling Toolbox for Computer Based Design. {\em Journal Of Machine Learning Research}. \textbf{11}, 2051-2055 (2010), http://jmlr.org/papers/v11/gorissen10a.html


\bibitem{FERUM}Bourinet, J., Mattrand, C. \& Dubourg, V. A review of recent features and improvements added to FERUM software. {\em Proc. Of The 10th International Conference On Structural Safety And Reliability (ICOSSAR’09)}. (2009)

\bibitem{UncertaintyPy}Tennøe, S., Halnes, G. \& Einevoll, G. Uncertainpy: A Python Toolbox for Uncertainty Quantification and Sensitivity Analysis in Computational Neuroscience. {\em Frontiers In Neuroinformatics}. \textbf{12} (2018), https://www.frontiersin.org/articles/10.3389/fninf.2018.00049

\bibitem{PyROM}Puzyrev, V., Ghommem, M. \& Meka, S. pyROM: A computational framework for reduced order modeling. {\em Journal Of Computational Science}. \textbf{30} pp. 157-173 (2019), https://www.sciencedirect.com/science/article/pii/S1877750318307518

\bibitem{Chaospy}Feinberg, J. \& Langtangen, H. Chaospy: An open source tool for designing methods of uncertainty quantification. {\em Journal Of Computational Science}. \textbf{11} pp. 46-57 (2015), https://www.sciencedirect.com/science/article/pii/S1877750315300119

\bibitem{SPUX}Šukys, J. \& Kattwinkel, M. SPUX: Scalable Particle Markov Chain Monte Carlo for uncertainty quantification in stochastic ecological models. (arXiv,2017), https://arxiv.org/abs/1711.01410

\bibitem{ABCpy}Dutta, R., Schoengens, M., Pacchiardi, L., Ummadisingu, A., Widmer, N., Onnela, J. \& Mira, A. ABCpy: A high-performance computing perspective to approximate Bayesian computation. {\em ArXiv Preprint ArXiv:1711.04694}. (2017)

\bibitem{SALib}Herman, J. \& Usher, W. SALib: An open-source Python library for Sensitivity Analysis. {\em Journal Of Open Source Software}. \textbf{2}, 97 (2017), https://doi.org/10.21105/joss.00097

\bibitem{PyApprox}Jakeman, J. PyApprox: Enabling efficient model analysis..  (2022,8), https://www.osti.gov/biblio/1879614

\bibitem{NeuralUQ}Zou, Z., Meng, X., Psaros, A. \& Karniadakis, G. NeuralUQ: A comprehensive library for uncertainty quantification in neural differential equations and operators. (arXiv,2022), https://arxiv.org/abs/2208.11866

\bibitem{sphinx_gallery}Nájera, Ó., Larson, E., Estève, L., Varoquaux, G., Liu, L., Grobler, J., Andrade, E., Holdgraf, C., Gramfort, A., Jas, M., Nothman, J., Grisel, O., Varoquaux, N., Gouillart, E., Luessi, M., Lee, A., Vanderplas, J., Hoffmann, T., Caswell, T., Sullivan, B., Batula, A., Jaeilepp, Robitaille, T., Appelhoff, S., Kunzmann, P., Geier, M., Lars, Sunden, K., Stańczak, D. \& Shih, A. sphinx-gallery/sphinx-gallery: Release v0.7.0. (Zenodo,2020,5), https://doi.org/10.5281/zenodo.3838216

\bibitem{PyMC}Salvatier, J., Wiecki, T. \& Fonnesbeck, C. Probabilistic programming in Python using PyMC3. {\em PeerJ Computer Science}. \textbf{2} pp. e55 (2016,4), https://doi.org/10.7717/peerj-cs.55

\bibitem{PyGPC}Weise, K., Poßner, L., Müller, E., Gast, R. \& Knösche, T. Pygpc: A sensitivity and uncertainty analysis toolbox for Python. {\em SoftwareX}. \textbf{11} pp. 100450 (2020), https://www.sciencedirect.com/science/article/pii/S2352711020300078

\bibitem{Fortuna}Detommaso, G., Gasparin, A., Donini, M., Seeger, M., Wilson, A. \& Archambeau, C. Fortuna: A Library for Uncertainty Quantification in Deep Learning. {\em ArXiv Preprint ArXiv:2302.04019}. (2023)



\end{thebibliography}


\section*{Current executable software version}\label{CurrentExec}

\begin{table}[!h]
\begin{tabular}{|l|p{6.5cm}|p{7.5cm}|}
\hline
\textbf{Nr.} & \textbf{(Executable) software metadata description} & \textbf{Please fill in this column} \\
\hline
S1 & Current software version & 4.1.0 \\
\hline
S2 & Permanent link to executables of this version  & 
\url{pypi.org/project/UQpy}

\url{github.com/SURGroup/UQpy/releases} 
\url{anaconda.org/conda-forge/uqpy} \\
\hline
S3 & Legal Software License & MIT License \\
\hline
S4 & Computing platforms/Operating Systems & Linux, OS X, Microsoft Windows \\
\hline
S5 & Installation requirements \& dependencies & see \texttt{setup.py}\\
\hline
S6 & If available, link to user manual - if formally published include a reference to the publication in the reference list & \url{https://uqpyproject.readthedocs.io} \\
\hline
\end{tabular}
\caption{Software metadata}
\label{sec:SoftwareMetadata} 
\end{table}

\end{document}